\documentclass[12pt,preprint]{article}
\usepackage{graphicx}
\usepackage{journals}
\usepackage{natbib}

\begin{document}
\begin{center}
{\Huge\bfseries Galactic kU-band Thermal Survey (GUTS)\footnote{White Paper submitted for the NRAO VLA Sky Survey\\ \hspace*{2cm}https://science.nrao.edu/science/surveys/vlass/call-for-vlass-white-papers}}\bigskip

{\large\bfseries Lor\'ant O.\ Sjouwerman \& Elisabeth A.C.\ Mills}\medskip

National Radio Astronomy Observatory, Socorro, New Mexico, USA \\
email: {\tt lsjouwer@nrao.edu, bmills@nrao.edu}

{\itshape December 14, 2013}
\end{center}

\begin{abstract}

This White Paper proposes a unique milli-Jansky sensitive arcsecond
resolution VLA 15 GHz (Ku band) continuum polarization imaging survey
of the Galactic plane using C array configuration. The science driver
is to fill the gap between previous lower frequency VLA Galactic plane
surveys, mainly detecting synchrotron radio emission at 1.4 and 5 GHz,
and the higher frequency (future ALMA and present {\itshape MSX,
  SPITZER, PLANCK, WISE}) 0.1-100 THz infrared surveys that detect
thermal radiation. The 15 GHz band appears to be the highest frequency
at which such a survey is both feasible and maximally unique from the
synchrotron radio view and will facilitate new multi-waveband Galactic
research and, for example, identify targets for ALMA. Depending on the
final survey parameters, for example a 3 second integration time for a
(8$\sigma$) 1 mJy 12-18 GHz detection limit over a 10 degree band
along the VLA observable (280 degree) Galactic plane, this survey
would consume up to 3200 VLA observing hours using on-the-fly mapping
and still allow for, e.g., an additional sub-Jy sensitive 12.2 GHz
methanol maser line survey. Furthermore this survey would reveal
potential compact high-frequency VLA calibrators in the Galactic bulge
and plane, a region where there is a severe lack of them.

\end{abstract}

\section*{Motivation}

Whether they simply cover large areas on the sky or only a very
detailed view of a specific targeted region of high interest, the
scientific yield of surveys is immensely large, with diverse
applications, cross-waveband value and legacy importance.

To be outlined here, albeit also motivated by our own research area,
is a survey feasible with the VLA and available data reduction
techniques, that in our opinion would make the largest impact possible
in a maximum number of astrophysical fields.

With a typical division of surveys into deep-field and/or large-sky
extragalactic surveys and surveys of (select areas of) the Milky Way
galaxy, we are most familiar with the latter and thus here we will
concentrate on a radio survey of the Milky Way galaxy.

Given the impact of an instantaneous large field of view on survey
speed, it is tempting to opt for surveys at the lower frequency bands
for the VLA. Indeed some of these surveys have been done with the VLA
prior to the upgrade, such as the {\sf NVSS, FIRST, CORNISH} and {\sf
  VGPS} surveys \citep{nvss98,fir95,corn12,vgps06}. However,
regardless of the scientific motivation for these previous surveys,
the lower frequencies ($\nu \leq 5$\,GHz) predominantly detect sources
of non-thermal synchrotron radiation and are not very well suited to
study thermal sources that radiate more of their energy at higher
radio frequencies ($\nu \geq 10$\,GHz).

Given the recent wealth of detailed large sky-area
arcsecond-resolution observations with sophisticated spacecraft in the
infrared ({\itshape IRAS, MSX, AKARI, SPITZER, HERSCHEL, WISE, PLANCK,
  ...}), at wavebands that are not hindered by interstellar extinction
(just like in the radio regime), it seems obvious to, as a first step,
match the (thermal) radiation from dust, star forming regions,
planetary nebulae, etc., detected in these surveys with their radio
counterparts. However, without disregarding the surveys done with
other instruments at higher frequencies, and future surveys to be done
at even higher radio frequencies ($\nu \geq 100$\,GHz) with ALMA (when
it becomes acceptive to large-area surveys), so far a milli-Jansky
sensitive arcsecond resolution continuum polarization imaging survey
of the Galactic plane at a higher frequency ($\nu > 5$\,GHz) is
lacking. The science drivers for this survey are nicely outlined,
among others, e.g., in the {\sf CORNISH} \citep{corn08,corn12} and
{\sf AMIGPS} \citep{ami13} papers.  Please also imagine inserting your
own favorite Galactic science topic here. We are sure there are many
that we have not thought of that may well matched to a survey like
presented here, with or without slight modifications (to be
discussed).

Here we give some arguments why a group of select and similar previous
continuum surveys are insufficient for the multi-wavelength comparison
studies that we argue should be a driving motivation of the VLA Sky
Survey (VLASS). Below we also discuss some line options, but line
sensitivity has a large impact on the request for observing time (for
a specific coverage on the sky) and are thus only realistically
feasible for select areas of interest. While covering such specific
target areas may be of high importance for some fields in astrophysics
(insert your favorite target here), we choose to focus on the more
general applications of larger scale continuum surveys.

\begin{itemize}

\item{\sf\bfseries AT20G\ }\citep{atk10,atk11} This all-sky ATCA
  survey at 20 GHz is sensitive to non-synchrotron emission and covers
  the sky below Dec.\ 0$^\circ$. Though the survey in principle covers
  the inner Galaxy and Southern plane, it unfortunately omits the
  Galactic plane itself (covering $|b| > 1.5^\circ$) and lacks
  coverage in the well studied fringes of the Northern part of the
  plane ($0^\circ < {\textrm Dec.} < +40^\circ$) also accessible with
  the ATCA. Furthermore, in comparison to infrared surveys like {\sf
    GLIMPSE} the angular resolution is poor
  ($\sim$100$^{\prime\prime}$). At an RMS noise level of about 10
  mJy/beam (detection limit just under 100 mJy/beam), {\sf AT20G} is
  sufficiently sensitive to detect high-frequency calibrators for the
  pre-expanded VLA at the higher Galactic latitudes.

\item{\sf\bfseries GPA\ }\citep{gpa00} This survey of almost the
  entire Northern sky Galactic plane ($-15^\circ < l < 255^\circ$)
  also covers a considerable range of Galactic latitude ($|b|<
  5^\circ$), and therefore almost completely covers the {\sf MSX}
  infrared survey visible from the Northern Hemisphere. The observing
  frequencies of $\sim$8 and $\sim$14 GHz in principle would detect
  non-synchrotron sources, but the RMS sensitivity of $\sim$230 and
  $\sim$800 mJy/beam, respectively, is poor as well as the angular
  resolution ($\sim$580$^{\prime\prime}$ and
  $\sim$400$^{\prime\prime}$) of these single dish
  observations. However, these low resolution maps may help augment
  sensitivity to large scale structure not detectable in VLA snapshots
  or with on-the-fly (OTF) mapping in the VLASS.

\item{\sf\bfseries AMIGPS\ }\citep{ami13} An interferometric survey at
  $\sim$16 GHz that has more than a factor of two better angular
  resolution, and almost a factor of 300 better noise ($\sim$3
  mJy/beam) compared to the single dish {\sf GPA} survey over the same
  Galactic latitude ($|b|< 5^\circ$). Unfortunately this survey only
  covers regions of the Galactic plane that pass at high elevations
  ($76^\circ < l < 170^\circ$), which critically will still avoid the
  Galactic center and bulge even after the survey is extended to lower
  elevations in the future ($53^\circ < l < 193^\circ$).

\item{\sf\bfseries VLA-L\ }\citep{vlal90} To our knowledge this is the
  Galactic plane survey with the largest range in Galactic longitude
  ($-20^\circ < l < 120^\circ$) using the (pre-expanded) VLA at 1.4
  GHz. The Galactic latitude is fairly limited to a single pointing on
  each side of the Galactic equator ($|b|< 0.8^\circ$), but the
  angular resolution is matching ($\sim$5$^{\prime\prime}$) for
  comparison with most infrared catalogs and images. The RMS of
  $\sim$10 mJy/beam is reasonable, but the largest problem for the
  study of thermal or non-synchrotron radio sources is that this
  survey is performed at the low frequency of 1.4 GHz.

\item{\sf\bfseries VLA-C\ }\citep{vlac94} This 5 GHz survey is similar
  to the {\sf VLA-L} survey with a fairly limited Galactic latitude of
  a single pointing on each side of the Galactic equator ($|b|<
  0.4^\circ$) and also a 2/3 cut in longitude coverage ($-10^\circ < l
  < 40^\circ$). However, observing the Galactic plane at a three-times
  higher frequency (4.9 GHz) it also has an angular resolution
  ($\sim$4$^{\prime\prime}$) well matched to infrared surveys and
  generally a better RMS (2.5-10 mJy/beam). This makes the combination
  of the {\sf VLA-L} and {\sf VLA-C} surveys an excellent resource to
  obtain spectral indexes of radio sources in the 40 square-degree
  overlap region and determine whether sources are non-synchrotron
  emitters.

\item{\sf\bfseries CORNISH\ }\citep{corn12} To our knowledge this is
  the Galactic plane survey with the highest angular resolution
  ($\sim$1$^{\prime\prime}$) and the best sensitivity (0.4 mJy/beam
  RMS). Observing at a frequency of 5 GHz, it was designed to match
  the {\itshape SPITZER} {\sf GLIMPSE} infrared survey of part of the
  Galactic plane, North of the Galactic bulge, with a similar angular
  resolution ($|b|< 0.8^\circ$, $10^\circ < l < 65^\circ$). The main
  science driver was to detect UCH{\sf II} regions produced by B3
  stars across the Galaxy, as well as detecting, among other sources,
  planetary nebulae, ionized winds from evolved massive stars, active
  stars, binaries, non-thermal emission from active stars, high energy
  sources, active galactic nuclei and radio galaxies. The
  investigators also point out the legacy applications beyond the
  original scope and drivers. Unfortunately, this survey omits the
  Galactic bulge and center, and higher Galactic latitudes where many
  of these sources will be found as well. Furthermore, as this survey
  was done at a low frequency (5 GHz), it is still primarily sensitive
  to synchrotron radiation.

\item{\sf\bfseries MeerGAL\ }(Mark
  Thompson\footnote{\hspace*{-0.03in}pysalt.salt.ac.za/meetings/talks\_saao\_meerkat/session3/SALT-MeerKAT\%20\%20MeerGAL.pptx
    www.herts.ac.uk/research/stri/research-areas/car/surveys/surveys-of-the-milky-way/radio-surveys}
  \& Sharmila Goedhart) is a planned sensitive (better than 0.1
  mJy/beam) Southern Galactic plane survey ($|b|< 1^\circ$, $280^\circ
  < l < 350^\circ$) with MeerKAT to start after 2018 in the 10-14 GHz
  frequency range. We note that this longitude range almost completely
  covers the longitude range unavailable to the VLA. Details about
  angular resolution ($\sim$0.2$^{\prime\prime}$?) are not readily
  available, but we imagine that MeerGAL and the here proposed survey
  are completely complementary.

\end{itemize}

With the exception of {\sf MeerGAL}, which has yet to be executed, the
above surveys were designed to achieve specific science goals and have
done just that very well. However, to characterize the emission from
the majority of components in the Galactic plane, bulge and center,
which involves stars in any stage of their evolution and warm and
dusty cloud cores, these surveys are too biased toward synchrotron
radiation, or toward extragalactic AGN. To address the shortcomings of
the above surveys, a new VLA Sky Survey of the Milky Way should aim to
cover as much of the Galactic plane, bulge and center (as large as
currently pragmatic), at a frequency higher than previous surveys (as
high as currently pragmatic), at the best possible sensitivity (as
sensitive as currently pragmatic) and at an angular resolution that is
well-matched to the best angular resolution ($\leq$1$^{\prime\prime}$)
of recent infrared surveys.  Essentially a VLA survey, at a (high)
frequency not used before, yielding a detailed view over a large area
of general interest (the Galactic center, bulge and plane) for which
data in other wavebands (on either side of the spectrum) is readily
available for comparison and spectral index measurements. Such a new
VLA Sky Survey would yield the least duplication of previous work and
thus make the highest possible impact in astrophysics at this stage,
and as a bonus, dramatically increase the density of high-frequency
calibrator sources in a region in the sky where they are highly
necessary but still lacking.

\bigskip

That is, we propose {\bfseries a VLA survey that combines the superior
  sensitivity and angular resolution of the {\sf\bfseries CORNISH}
  survey, with the sky area covered by the low resolution
  {\sf\bfseries GPA} survey, while observing at a frequency comparable
  to the {\sf\bfseries AT20G} survey}, a survey that, given the
capabilities of the VLA, is long overdue.

\section*{Observing Parameters}


To detect radio emission from non-synchrotron (thermal) sources the
best choice is a radio frequency at the higher end of the receiver
ranges. Unfortunately, the original VLA was not built with this in
mind, meaning that observations at the highest end of the receiver
range ($\nu >$ 20 GHz) require significant extra overhead of antenna
pointing. Furthermore, the antenna dishes are only sufficiently large
to yield fields of view of the order of an arcminute or two,
significantly affecting the survey speed. The highest frequency that
may avoid the extra overhead of pointing seems to be the 12-18 GHz
observing band (Ku, also known as U or 15 GHz, receivers). With 6 GHz
continuum bandwidth this range is as sensitive as the 8-12 GHz band
(X) and, due to the effect of the atmosphere, more sensitive than any
of the higher frequency (K, Ka, Q) bands for which one can cover a
full 8 GHz bandwidth. The center frequency of 15 GHz also is a natural
``factor-3'' extension of pre-expanded VLA surveys at 1.4 and 5
GHz. The next factor-3 frequency would be 45 GHz, also available at
the VLA, but probably the topic of a next generation VLASS.

At 15 GHz, to obtain the $\sim$1$^{\prime\prime}$ angular resolution
similar to that of the 5 GHz {\sf CORNISH} survey, the natural choice
for array configuration would be C, or CnB for the lower
Declinations. Perhaps the D and DnC would better match the
few-arcsecond angular resolution of the {\sf VLA-L} and {\sf VLA-C}
surveys, but we would argue that one would want to match the best
angular resolution available in the infrared while making use of an
array configuration that is less oversubscribed. This is also a
benefit for detecting the most compact sources that can be used for
phase calibration at the higher frequencies. The largest angular scale
in snapshot observations is just smaller than an arcminute for any of
these arrays, which is sufficient for comparison with infrared
maps. Detecting gas (apart from masers) is not an essential part for
this survey and therefore does not impose the constraint to observe in
the smallest (D, DnC) array configurations. Note that the same
$\sim$1$^{\prime\prime}$ angular resolution is obtained at 45 GHz with
the D array configuration.

While the survey will be set up as a continuum polarization survey,
which only has a minimal effect on total calibration overhead,
correlator resources are available to include searches for narrower
line emission other than with the default 2 MHz ($\sim$40
km$\cdot$s$^{-1}$) channel separations, at the cost of a higher data
rate and data volume and a slightly more complicated data
reduction. The details will vary with the details of the line setup,
but the point of this ``almost for free'' extra survey data is that it
should be included.

The catalog compiled by Frank
Lovas\footnote{http://physics.nist.gov/cgi-bin/micro/table5/start.pl}
lists about 80 lines detected previously in what can be considered the
VLA Ku-band frequency range (11.9-18.1 GHz). Probably the most
interesting categories are those of masers (CH$_3$OH, OH, maybe SiS),
thermal cores (H$_2$CO, CH$_3$OH, maybe NH$_3$) and the vast amount of
carbon molecules, such as variations of HCN, HC$_x$N (with $x$ an odd
positive integer $<$10). Easiest to observe are the (methanol) masers
as they typically are compact (point-like in any array configuration)
and bright, and can be detected in a minimum scan time over the entire
Galaxy, but it won't hurt to include others if resources are
remaining. The X band range (7.9-12.1) returns about 65 lines of
mostly other transitions of the same species.

\begin{table*}[t]\caption{On-the-fly observations at 3 sec field
    integrations}{\smallskip} 
\begin{tabular}{ccccrrrcr}
\hline\\
$\nu$&FOV&BW& RMS &slew&10$^\circ$ strip&$\Delta l$=280$^\circ$&Time&Total\\
(GHz)&($^\prime$)&(GHz)&($^a$)&($^{\prime\prime}$/s)&(min)&\# strips&(hours)&(hours)\\
{\smallskip}\\
\hline\\
18-26/K  &1.23 &7.5& 225 & 25 &24+oh&13650&5460+oh$^b$&$\sim$7000\\
12-18/Ku &1.78 &5.5& 130 & 36 &17+oh& 9450&2680+oh&$\sim$3200\\
8-12/X   &2.67 &3.5& 130 & 53 &11+oh& 6300&1155+oh&$\sim$1400\\
2-4/S    &11.3 &1.5& 210 & 160&3.8+oh&2100&133+oh&$\sim$150\\
\\
\end{tabular}
\hrule\smallskip
$^a$ RMS in $\mu$Jy/beam\\
$^b$ Includes extra overhead for pointing scans, next to standard
calibration for flux density, bandpass, delay/phase and polarization\\
\end{table*}


The intrinsic continuum sensitivity is impressive which means that for
a very reasonable detection limit only a few seconds of integration
are needed per field to achieve sub-mJy sensitivity (see
Table~1). With only a few seconds per pointing, it will be impractical
to use the individual field point-and-shoot approach as the slewing
overhead will be larger than the total on-source observing
time. Therefore for any such survey an on-the-fly (OTF) mapping mosaic
is the way to go. In Table~1 the RMS and survey numbers are given for
a 3 second field integration time for the 18-26, 12-18, 8-12 and 2-4
GHz ranges. The latter is included to show that indeed the large
primary beam (FOV) of the lower frequencies is attractive in covering
an area in the sky with minimum observing time, but that only holds if
the observing frequency is more flexible and not a key parameter in
the survey, as it must be to be most sensitive to thermal sources. The
former is included for the opposite reason, as not only does the time
per area become large at the higher frequencies, but for frequencies
over about 20 GHz a significant amount of overhead needs to be
included for antenna pointing scans. Also, it would impact the
observing time available during high-frequency weather conditions
considerably. The sweet spot in terms of sensitivity lies in the range
4-18 GHz, where we omitted to list the range 4-8 GHz (C band) as a
partial Galactic survey in this range has already been done ({\sf
  VLA-C} and {\sf CORNISH}). The preference of the 12-18 GHz range
over the 8-12 GHz range, which both are similar in RFI and other
observing effects and constraints, then becomes a matter of potential
return. 
Our preference for the higher frequencies is motivated by:\\
(-) the relative increase of non-synchrotron versus synchrotron
radiation,\\
(-) the ``factor-3'' with respect to the previous VLA surveys,\\
(-) the higher resolution,\\
(-) the inclusion of the strong methanol maser at 12.2 GHz,\\
(-) a frequency that matches the missing Galactic plane data for the
{\sf AT20G} survey,\\
(-) a previously largely unused frequency range, and\\
(-) a new homogeneous survey that would not be sidetracked by
re-observing or skipping the (non-homogeneous) areas from previous
observations.\\

In Table~1 we have compiled the time it would take to survey the
Galaxy for $|b|<5^\circ, -20^\circ < l < 260^\circ$ (2800 square
degrees), i.e. the {\sf GPA} and {\sf MSX} survey area expanded to or
overlapping with the maximum coverage at the VLA (that can reach
slightly lower Declinations than the GBT). It also shows the time it
takes for a 10$^\circ$ strip (i.e., sampling in $b$) and the number of
such strips to cover the 280$^\circ$ of the entire visible Galactic
plane (i.e. $l$) with the slew rate for 3 seconds per field. The raw
and approximate (i.e., including overhead) total time for the whole
survey is also listed for an integration time of 3 seconds per
field. The integration time of 3 seconds is used as this would give an
8$\sigma$ continuum detection limit of 1 mJy. This is comparable (but
slightly better) to the sensitivity of the {\sf CORNISH} survey and
results, with 1 second dump times, in a data rate of 36 MB/s (128
GB/h) and a total data volume of 350 TB (excluding extra resources
used for lines) when observing a 6 GHz bandwidth at 15 GHz. The
numbers are a factor 1.5 smaller for the 4 GHz bandwidth at 10 GHz.

Note that with a detection limit of 1 mJy, continuum calibrators as
faint as 25 mJy (8$\sigma$ on a single polarization baseline, both at
10 and 15 GHz) will be detected and greatly expand the list of
calibrators available in and near the Galactic plane compared to the
current cut-off of 100 mJy.


\subsection*{Possible alternatives}

The survey parameters described above are designed to yield the
maximum scientific return for a Galactic VLASS.  Although it is
technically feasible (and indeed scientifically desirable) with the
current observational status of the VLA to perform this survey, it is
painful to think of alternatives for descoping if the survey would be
considered {\slshape logistically} not feasible, even when spread out
over many configuration cycles. There are however some alternatives,
each with their trade-off consequences:

\begin{itemize}

\item{\sf\bfseries Slew rate/integration time/sensitivity:\ }Slew rate
  and integration time for a single epoch observation are both coupled
  such that changing either parameter impacts the overall
  sensitivity. We would point out that, e.g., doubling the slew rate
  or halving the observing time for a {\small $\sqrt(2)$} cut in
  sensitivity, would mean increasing the dump rate by at least a
  factor two and thus increasing the data rate and data volumes by at
  least a factor of two. This may cause operational problems, but is
  certainly a point of consideration.

\item{\sf\bfseries Survey area:\ }Cutting the survey area will cut the
  total observing time linearly with the coverage. One could argue
  that the outer Galaxy or the higher latitudes are less
  interesting. However, for legacy value and for coverage overlap with
  the {\sf MSX} and {\sf GPA} surveys, as well as for currently
  neglected studies in the outer Galaxy and higher latitudes, e.g.,
  where the metallicity of the objects is much lower than for the
  inner Galaxy, it is important to maintain the goal of observing the
  fully overlapping accessible survey area.

\item{\sf\bfseries Overhead:\ }The only luxury overhead included is
  the overhead for polarization calibration. However, the overhead is
  minimal and the extra scientific return of the availability of
  polarization measurements tremendously outweigh the small increase
  in observing time.

\item{\sf\bfseries Frequency:\ }Observing at a lower frequency, like
  in the 4-8 GHz (C band) range, would speed up the survey speed due
  to the larger primary beam. However, observing at a lower frequency
  would severely impact the science goal of surveying non-synchrotron
  radiation.

\item{\sf\bfseries Upscoping!\ } Perform this survey in the 40-48 GHz
  range (Q band) in D array configuration. The scientific return for
  thermal sources, a diversity of (maser) lines and high-frequency
  continuum calibrators would be much larger, but at a cost of a
  hugely increased demand on high-frequency weather observing time
  (and thus probably not feasible).

\end{itemize}

\vfill
\noindent In summary: No GUTS, no glory!
\vfill



\end{document}